\documentclass[aps,pre,twocolumn,superscriptaddr+ess,showpacs,showkeys]{revtex4-2}

\usepackage{graphicx}
\usepackage{dcolumn}
\usepackage{bm}
\usepackage{enumitem}
\setlist[itemize]{noitemsep, topsep=0pt}
\setlist[itemize]{leftmargin=*}

\begin{document}

\preprint{APS/PRE}

\title{Capillary Pressure-Saturation relation derived from the Pore Morphology Method}

\author{Fernando Alonso-Marroquin}
 \email{fernando.marroquin@kfupm.edu.sa}
\author{Martin P. Andersson}%

\affiliation{CIPR, King Fahd University of Petroleum and Minerals, Dhahran, 31261, Kingdom of Saudi Arabia. }

\date{\today}

\begin{abstract}
A computationally efficient method to calculate the capillary pressure-saturation relations of immiscible multiphase flow on two-dimensional pore morphologies is presented here. The method is an extension of the Pore Morphology Method that includes wetting angle and trapped mechanism of the displaced fluid, and calculation of material properties by density functional theory. After validating the method with micro-chip fluid injection experiments, the method is used to relate pore morphology to capillary pressure-saturation relation using square-lattice pore morphologies. Because the method uses only morphological binary operations, it is more efficient than well-established high-resolution voxel dynamics methods such as Lattice Boltzmann Methods and Level-set computational fluid dynamics. Apart from pore morphology, only the material parameters related to contact angle (wettability) and interfacial tension are required to connect the pore-saturation relation and pore throat distribution. We investigate the effect on interfacial tension, wettability, sample size, and pore throat distribution on entry pressure and residual saturation.
\end{abstract}


\maketitle

\section{Introduction}
The displacement of a fluid by another in porous media such as soils, rocks, and engineering materials is a complex process, difficult to tackle from first principles. A common simplification is to assume that fluids are immiscible and incompressible, so that the process is governed by the pore morphology and fluid properties such as interfacial tension, wettability, and viscosity \cite{blunt2017multiphase}. The original Lucas-Washburn model set the mathematical basis of the process \cite{washburn1921dynamics}, and closed-form solutions are available for uniform cylindrical geometries, accounting for effects of dynamics \cite{maggi2012multiphase}, temperature \cite{maggi2013temperature}, and multi-components \cite{maggi2012multicomponent}. For more complex geometries, first-principles solutions are not available in analytical forms.  The pressure-saturation relation is often used to characterize the invasion process; initially, the porous medium is fully saturated by a resident fluid, then an invasive fluid is injected and the saturation of the invaded fluid is measured. This process is performed in quasistatic fashion, thus, it is expected that viscous forces play a minor role, and the invasion process is governed by capillary forces at the fluid interfaces. In addition to it's own importance in understanding the fluid flow through a porous medium, the capillary pressure-saturation relation is also one of the key inputs in reservoir scale simulation, and therefore efficient simulation of pressure-saturation relation can also serve as a key component in scaling up molecular and pore scale phenomena to the field scale, via reservoir simulations.

Due to the exceptional growth of the computational power, fluid invasion is often simulated using computationally intensive voxel dynamics methods, such as the lattice Boltzmann method \cite{li2013lattice,premnath2005lattice}, and the level-set computational fluid dynamics \cite{yiotis2021pore}. Less attention has been paid to more efficient quasi-static methods, such as the Pore Morphology Method, where the quasi-static invasion is modeled by simpler morphological operations \cite{hilpert2001pore}. Voxel simulations are computationally expensive, a single simulation on $1000^3$ voxels requires millions of time-steps to be run on supercomputers. Binary algorithms are more favorable than voxel dynamics ones, since they do not pose time-steps instabilities and they involve much less computational operations.  

In this paper, we exploit the advantages of the Pore Morphology Method to calculate the quasi-static response of the saturation by the increase of capillary pressure in an extremely efficient way. Our starting point for the original Hilper and Miller algorithm \cite{hilpert2001pore}. In their method, the saturation at a given pressure is calculated by performing mathematical morphology operations and connectivity functions on the binarized pore image. Here we extend the method to account for trapped mechanism: resident fluid that gets disconnected from the evacuation zone is considered as trapped, so that the invasion is forbidden there. The extended Pore Morphology Method is presented in Section~\ref{sec: PMM}. The material constants involved in the simulations are given in terms of interfacial tensions that are calculated using the density functional theory in Section~\ref{sec:DFT}. The method is validated by comparison to two-dimensional microchip experiments in Section \ref{sec:microchip}. In Section~\ref{sec:simulations} we present the dimensional analysis of the pressure-saturation relation and perform numerical simulations on square lattices to investigate the effects of material and geometric heterogeneity of the pressure-saturation relations. We also investigate the finite-size effects to determine whether the pressure-saturation relations hold in the statistical limit where the sample is large enough to become a representative element for large-scale reservoir simulation.  Sections~\ref{sec:discussion} and \ref{sec:conclusions} are devoted to the discussion of the results and the overall conclusions. 

\section{Pore Morphology Method}
\label{sec: PMM}

\begin{figure*}[t]
    \centering
    \includegraphics[trim={2.5cm 3cm 2.5cm 2cm},clip,width = 0.4\linewidth]{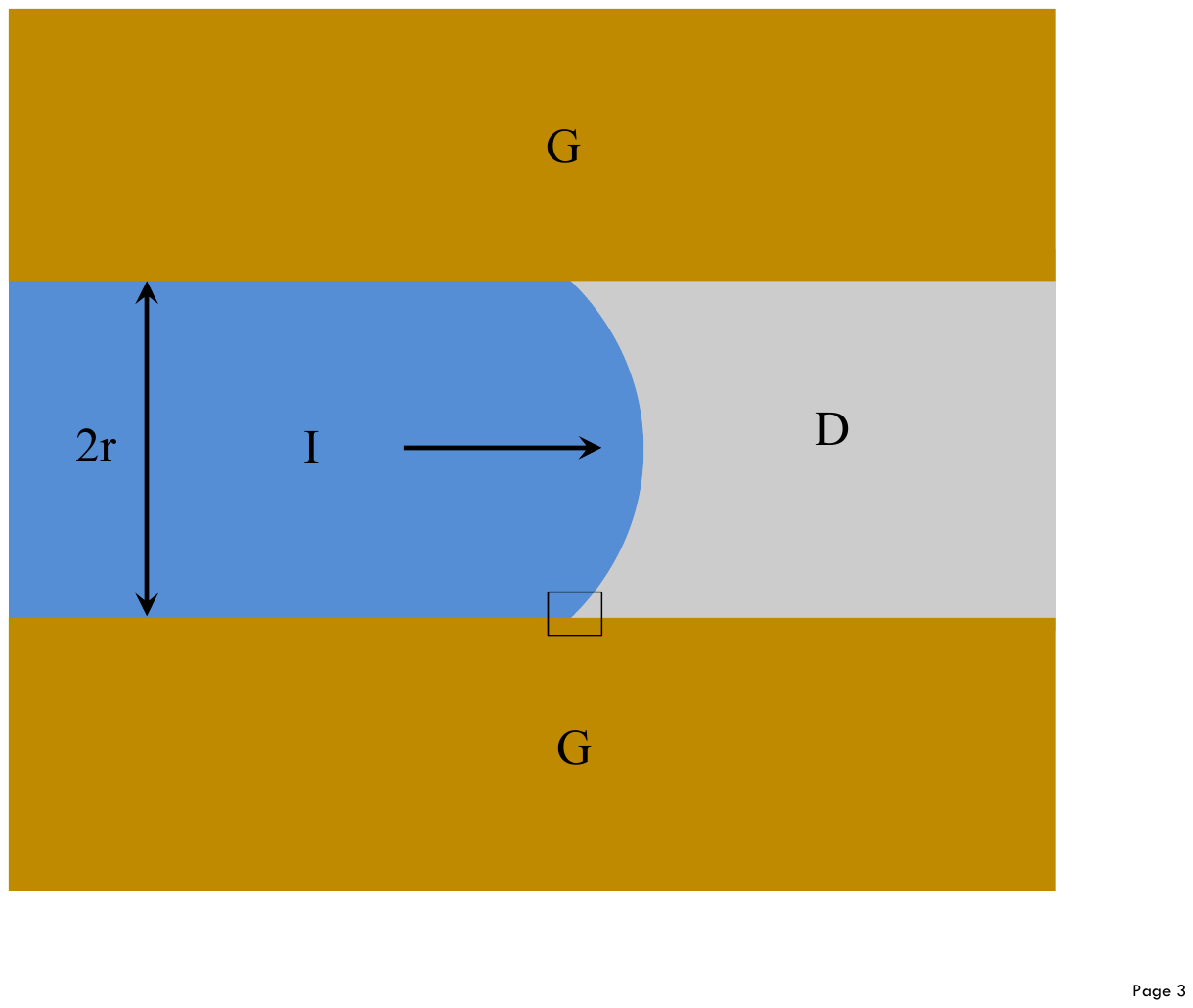}
    \includegraphics[trim={2.5cm 3.5cm 2.5cm 2.5cm},clip,width = 0.4\linewidth]{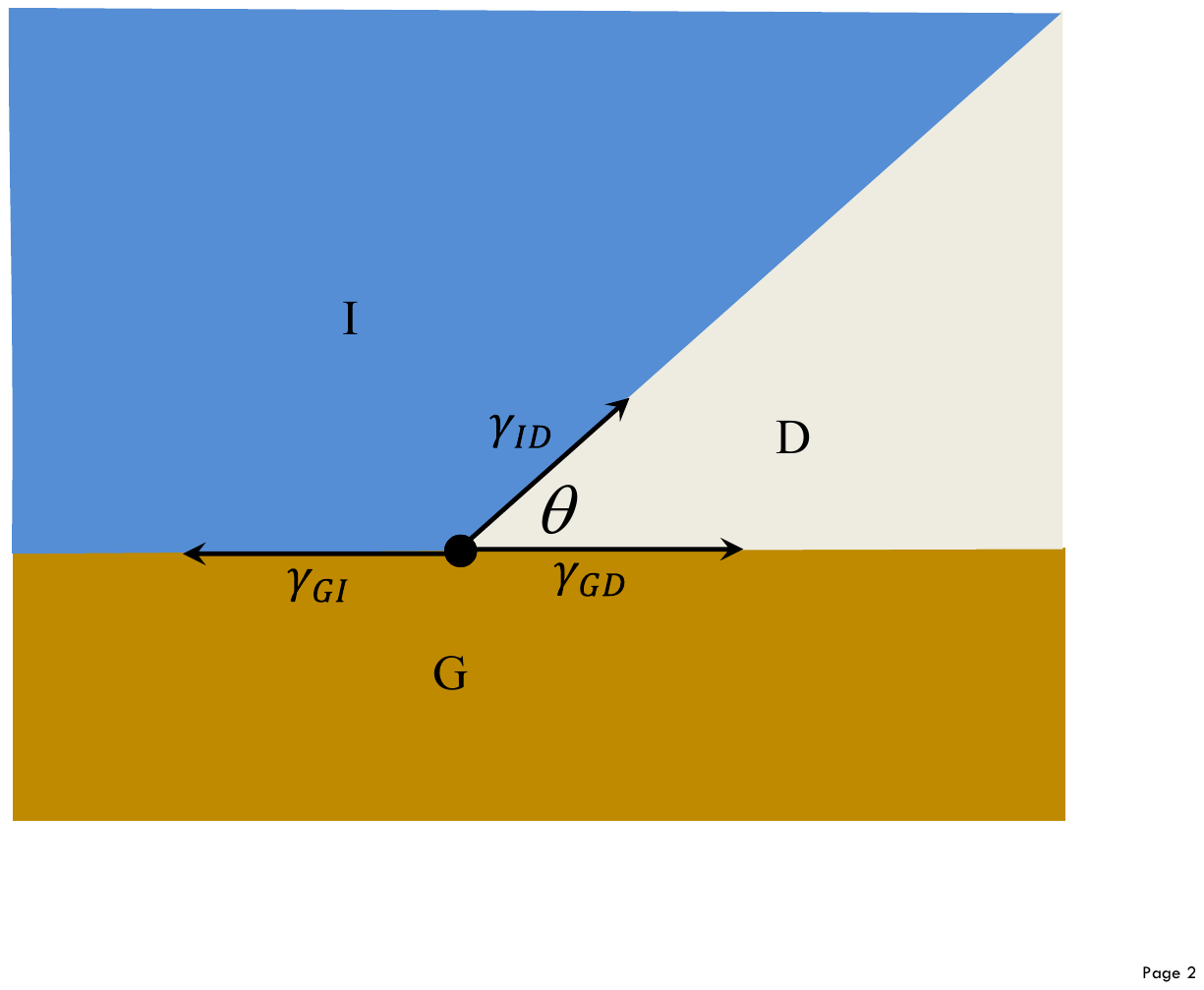}
    \caption{(a): Representation of a fluid (I) invading a second fluid (D) in a pore throat of radius $r$. The two phases intersect the solid granulate (G) into the contact line/curve. (b) Balance of linear tensions (force per unit of length), or equivalently interfacial free energies (energy per unit area) at the contact line.}
    \label{fig:Young_Laplace}
\end{figure*}

 Consider an incompressive resident fluid completely filling a porous medium. Then an invasive fluid is injected in a given region of the porous medium.  The medium is defined as a binary partition of a subset $S \subset \boldsymbol{R}^2$ of the Euclidean space into a granular matrix $G$ and a void space $V$; i. e.  $G \cup V = S $ and $G \cap V = \emptyset $. The void space is initially filled with the resident fluid that can escape through an {\it evacuation} zone $E \subset V$. The invading fluid is injected to the void space through a region $R \subset V$ that we called {\it reservoir}. The invasive fluid displaces the resident fluid in an immiscible fashion by the action of a capillary pressure $\Delta P_c$.  The {\it invaded} fluid is labeled by $I$ while the resident fluid is named as {\it defensive} fluid and labeled by $D$. The invasion process is governed by the equilibrium equation of the contact line/curve where the three phases intersect \cite{fan2020microscopic}, see Figure~\ref{fig:Young_Laplace}b
\begin{equation}
    \gamma_{GD} - \gamma_{GI} = \gamma_{ID}\cos\theta,
    \label{eq:Young}
\end{equation}
where $\gamma_{GD}$, $\gamma_{GI}$, and $\gamma_{ID}$ are the interfacial tension between granular-defensive fluid, granular-invasive fluid, and invasive-defensive fluids. the variable $\theta$ is the {\it contact angle}, often referred as wetting angle in the literature \cite{blunt2017multiphase}.
The free-body diagram of the contact interface is shown in Fig.~\ref{fig:Young_Laplace}b. Let us define $\Delta P_c$ as the capillary pressure at the fluid-fluid interface.  The capillary pressure threshold for invading a throat of radius $r$ is obtained by balancing the forces along the throat, see Fig.~\ref{fig:Young_Laplace}b. The force due to the fluid pressure is  $F=\Delta P_c \times area =  \Delta P_c 2rd$, where $d$ is the depth of the specimen. The tension due to the capillary forces is  $T =\gamma_{ID} \times perimeter\times \cos{\theta} =\gamma_{LD}(2r+d)\cos{\theta}$. Balancing these forces, ($F=T$)  leads to the Young-Laplace equation
\begin{equation}
    \Delta P_c = 2\gamma\cos\theta(\frac{1}{2r}+\frac{1}{d}),
    \label{eq:Young Laplace}
\end{equation}
Were $\gamma=\gamma_{ID}$ is introduced to simplify the notation. We also remove the contribution of the out-of-plane surface tension by defining the effective capillary pressure as $P_c=2\gamma\cos\theta/d+\Delta P_c$. Thus, the Young-Laplace equation given by Eq.~\ref{eq:Young Laplace} reduces to
\begin{equation}
    r = \frac{\gamma\cos\theta}{P_c}.
    \label{eq:capillary radius}
\end{equation}
For given capillary pressure $P_c$,~Eq.~(\ref{eq:capillary radius}) defines the radius of all throats that can be potentially invaded. This radius is defined here as the {\it capillary radius}, and it will be treated as the control parameter in our simulations.

The original algorithm of Hilper and Miller \cite{hilpert2001pore}   calculates the invaded pore space in terms of the capillary radius using connectivity functions and mathematical morphological operations on binary images. The Hilper and Miller algorithm does not have any trapped mechanism; For a capillary radius $r$, the zone reached by the invading fluid is calculated as $I_r = C(I \ominus \hat{S}_r,R)\oplus \hat{S}_r$ where $\hat {S _r} = S_r \cup S_{r-1}$, and $S_r$ is the digital sphere of radius $r$. The function $A=C(B,R)$ returns the connected component of the image $B$ that is connected to the image $R$ \cite{hilpert2001pore}. 
 
Here we present a modification of the algorithm that considers a trapped mechanism: any evacuating fluid that is completely surrounded by the invaded fluid is assumed to be trapped. In our algorithm, the void space $V$ is partitioned into a defended zone $D$, an invaded zone $I$, and a trapped zone $T$, initially the capillary radius is set to the largest throat neck of the void space, then the capillary radius is progressive reduced and the invaded zone is calculated. The trapped zone is added when any part of the defended zone become disconnected from the evacuation area. The initial conditions in the simulation are $D=V$, $I=T= \emptyset$. To update these zones, we use the same connectivity function $A=C(B,R)$ of the Hilper \& Miller algorithm. We define the complement of a subset $A\subset V$ of the void space as $A'=V-A$. Based on these functions, the steps of the simulation loop are:

\begin{itemize}[noitemsep]
    \item [1] The void image is morphologically eroded with a sphere $S_r$ of radius $r$, resulting in $I_{ero} = V \ominus S_r$; 
    \item [2] The zone that is not trapped and is connected to the reservoir is dilated: $I_{new} = C(I_{ero}\cap T',R) \oplus S_r$. This set is added to the invaded zone: $I = I \cup I_{new}$;
    \item [3] The defended area is calculated as the void space that is not invaded and is connected to the evacuation area $D= C(V \cap I',E)$; 
    \item [4] The new trapped zone is the void space that is neither in the defended zone nor in the invaded zone $T_{new} = V \cap D' \cap I'$. This area is added to the trapped zone: $T = T \cup T_{new}$;
    \item [5] The capillary radius is decreased  ($r\to r-1$), return to step 1 unless $r=0$.
\end{itemize}

The algorithm is an improved version of the Hilper \& Miller method that allows trapping mechanism, and it reduces to the non-trapped pore invasion method of the Hilpert and Miller algorithm when $E=V$. The algorithm is numerically more efficient than voxel-dynamics based solver such as the Lattice Boltzmann Method or the level set method. This is because the operations in the Pore Morphology method are binary rather than floating point calculations. Since the algorithm is quasi-static, it does not have time-step numerical error. Instead, the origin of the numerical error is on the pixel-size, (image resolution). The method does not account for viscous effects, but this is a point in our favor, as we want to focus on the capillary effects only.

The calculations of throat sizes is convenient to initialize the capillary radius, and to further accelerate the algorithm. To obtain the list of throat sizes, a pore network extraction is needed. This procedure is based on segmentation of the void space into pores; each pore is connected to the other by throats. Then the interconnected pores can be represented by a pore network, or more precisely, my an undirected graph. The graph itself can be seen as an image where the pores are treated as super-pixels and throats are treated as connection between the super-pixels. The morphological operations and connectivity functions can be rewritten to operate on these graphs in more efficient code that is pixel-independent. Besides, the operations are  much lower since the number of super-pixels corresponds to the number of pores, a number that is much lower than the number of pixels of the binary image.

To extract the pore network, we perform a workflow algorithm that we call the {\it charting} of the pore space: First, we calculate two quantities from the binary image of $B$ the void space: the label matrix $L$ and the skeleton $\kappa$. The label image $L$ is calculated from $B$ by applying the standard watershed segmentation on the void space using the city-block distance transform. The skeleton $\kappa$ reduces the binary image into a 1-pixel wide curved lines, while preserving the topology and Euler number of the image. From the skeleton, we identify the branch points. These points are connecting different branches of the skeleton. The domains of $L$ that do not contain branch points of the skeleton are deemed as invalid pores and are labeled zero in the label matrix.  Following, a gray dilation algorithm is used to assign to the pixels of the invalid pores that are unresolved by the watershed segmentation. The techniques of this procedure are explained by Rabbani et. al. \cite{rabbani2014automated}. The charted pore space defined the pores as the domains of the final label matrix $L$, and the throats as the interfaces between the pore domains from where the throat radii are calculated.

\section{Molecular modeling}
\label{sec:DFT}

The material properties (contact angle and interfacial tension) are calculated using Density Functional Theory (DFT).
To demonstrate the potential of scaling up molecular modeling results to relevant pore-scale phenomena, we computed the liquid-liquid interfacial tension~\cite{Andersson2014} as well as the contact angle~\cite{Andersson2020} for our comparison to the microchip experiments~\cite{yiotis2021pore}. The experiments were conducted using a polydimethylsilane (PDMS) solid material, deionized water as invading fluid, and the polyfluorinated liquid FC43 as the defending fluid. We used Turbomole v7.8~\cite{Ahlrichs1989}, the BP functional~\cite{Becke1988}\cite{Perdew1986}, the TZVP basis set~\cite{Eichkorn1997} and COSMO implicit solvent model ~\cite{Klamt1993} for the DFT calculations of the surface energies of the constituents. All DFT calculations were conducted using COSMOtherm, the BP\_TZVP\_C30\_1601.ctd parameterization~\cite{Andersson2018} and the scripts detailed in ~\cite{Andersson2020}. We modeled the PDMS surface using a dimethylsilane pentamer, with the end monomers removed in the COSMOtherm calculations, a standard method for treating polymers using COSMO-RS. The results were averaged for weight factors with and without including the oxygens in the backbone of the PDMS molecule. This was done because the oxygens in the silane backbone are situated deeper from the surface and do not have full access to the fluid in contact. The methyl groups are located further out from the backbone and would be in contact with the fluid. Therefore, they were always included in the DFT calculations. The DFT calculations predicted the liquid-liquid interfacial tension between water and FC43 as well as the solid-liquid interfacial tensions between the PDMS solid surface and water and between PDMS and FC43. The contact angle was then calculated using Young's equation, Eq.~\ref{eq:Young}. Free surface energies are interpreted as interfacial tensions. The calculations of the free energies are presented in Table ~\ref{tab:dft}. The different assumptions on molecular interaction that were made in the calculations led to a large variation in the individual water-solid interfacial tensions, but the average value gave a contact angle in agreement with the curve fitting used in the experiments~\cite{yiotis2021pore}. We would interpret this in the sense that  water has a difficult time reaching the deep lying oxygen atoms of the backbone of the PDMS, and therefore, are only partially in contact with this part of the molecular surface. Thus, the average interfacial tension that includes both a molecular surface with and without the backbone oxygen atoms in the PDMS gives a result in agreement with experiments.  

\begin{table}[t]
    \centering
    \begin{tabular}{c|c|c|c} 
        \hline
        water-oil & water-solid & oil-solid  & weight factor \\ 
        \hline
        47.6  & 11.6  & 9.9 & excluding edge-bridging 0s  \\ 
              & 17.9  & 9.8 & including one edge-bridging 0 \\
              & 50.6  & 9.7 & including only CH3 and Si \\
              & 53.1  & 9.6 & including only CH3 groups \\
      average &       &     & \\
        47.6  & 33.3  & 9.7 & contact angle:  60$^o$ \\
        \hline
    \end{tabular}
    \caption{Density functional theory calculation of interfacial free energy (mN/m) using COSMO-RS. Oil corresponds to FC43; and solid to PDMS. The experimental water-oil surface tension is taken as $55$ mM/m and the experimental contact angle is curve-fitted as $\theta_0= (55\pm 5)^o$ by Yiotis et. al.~\cite{yiotis2021pore}}.
    \label{tab:dft}
\end{table}

\section{Comparison with microchip experiment}
\label{sec:microchip}

\begin{figure*}[t]
    \centering
    \includegraphics[trim={2.5cm 8cm 2.5cm 7cm},clip,width = \linewidth]{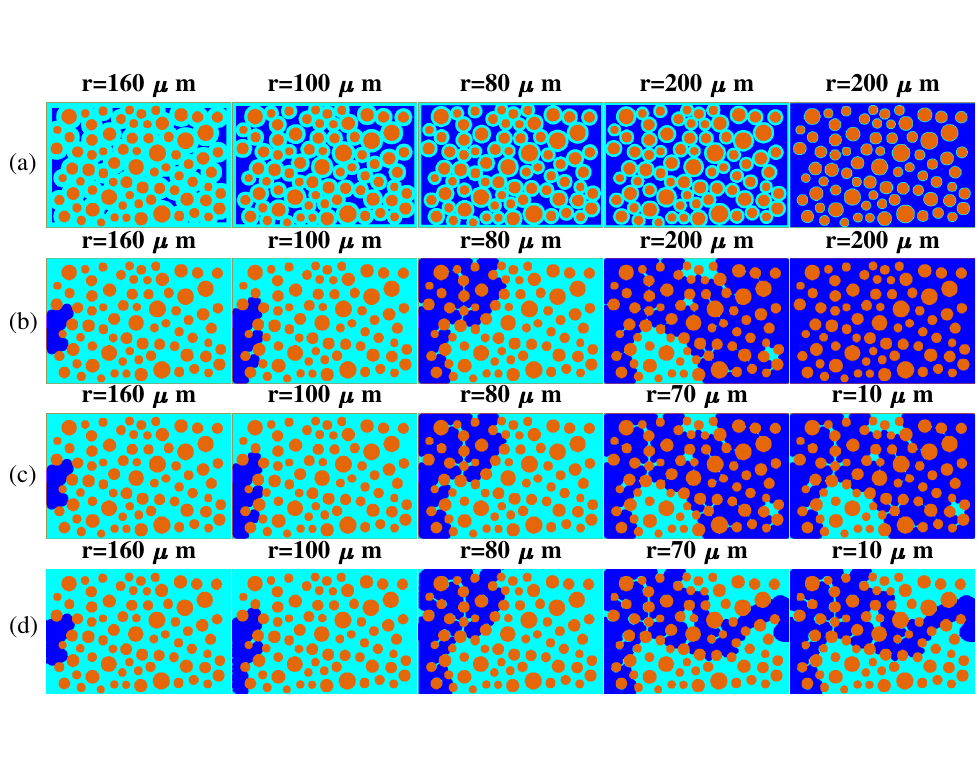}
    \caption{
    Snapshots from flow-controlled experiment and pressure-controlled simulations for different capillary radii. The capillary radius is given by $r = \gamma\cos\theta_0/P_c$ so that increasing capillary pressure leads to a decrease of the capillary radius. Snapshots in (a) show the morphological erosion of the void space by a disk or capilary radius $r$. Snapshots in (b) shows the pore morphology simulation results based on the original Hilper and Miller algorithm \cite{hilpert2001pore}, without trapping mechanism. In their algorithm, the zone invaded by the non-wetting fluid is calculated as the dilation of the eroded space in (a) connected to the reservoir. (c) shows the snapshots of our pore morphology simulation that accounts for trapped mechanism: defended (non-wetting) fluid that lose contact from the evacuation zone get trapped. (d) are the snapshot of the flow-rate controlled experiments at the peaks of the capillary pressure; notice that at the end of the experiment the pressure drops as the invasive fluid breaks through. The match between our simulation and experiments is perfect before the pressure peaks. Our pore morphology simulations does not capture the viscous fingering after pressure peaks, but they allows derivation of the pressure-saturation relation since the simulation is pressure-controlled and quasistatic.}
    \label{fig:mosaic}
\end{figure*}

To validate the numerical method, we compare to the two-dimensional experiment reported by Yiotis et. al. \cite{yiotis2021pore}. The micromodel consists of a rectangular area that is $6~ mm$ long and $4~ mm$ wide. It also includes two flow channels that are $13~ mm$ long and $500 ~\mu m$ wide connected to liquid buffers. The micromodel is $115 ~\mu m$ deep, and it contains $76$ non-overlapping cylindrical pillars with an average diameter of $380~ \mu m$. Oil is used as the wetting (defensive) fluid; and water as a non-wetting (invaded) fluid. The pressure sensor has a maximum pressure reading of $7000~ Pa$ and a typical precision of $20~ Pa$. Water is pumped on one side of the chip through a tube of $1~mm$ diameter, and the oil is evacuated on the other side through a tube of the same diameter. The water is injected with a constant flow rate of $10^{-4}~ ml/min$. The value of the interfacial tension reported in the experiment is $\gamma = 55~N/m$. The contact angle of $\theta_0 = 60^0$ was calculated based on Density Funcional Theory using the COSMO-RS software. The computed value gave very consistent results for the pressure-saturation relation. 

We capture the distribution of both wetting and non-wetting phase each time the pressure reaches a maximal value and a new pore is invaded. Then, the capillary radius is recorded as $r = \gamma\cos\theta_0/P_c$ at the point of maximal pressure. In the pore morphology simulation, the capillary radius is initialized with a value larger than the largest throat. Then in each simulation step the capillary radius is reduced by one pixel and the invaded and trapped areas are calculated using the pore morphology algorithm described above. The first step is to calculate the morphological erosion of the void space by a sphere of radius $r$, the capillary radius, as shown in the first line of Fig.~\ref{fig:mosaic}. In the original Hilper and Miller algorithm, the eroded space that is connected to the injection area is morphologically dilated, and the dilated space corresponds to the invaded zone, shown in the second line of Fig.~\ref{fig:mosaic}. The Hilper and Miller algorithm returns the saturation as a function of pressure, or capillary radius, but since there is no trapping mechanism for the defensive fluid, the saturation of the wetting phase reaches zero as the pressure increases (i.e, the capillary radius reaches zero). 

Our modified pore morphology method accounts for trapped of the wetting fluid by adding and additional rule: in each step, the defending fluid that get disconnected from the evacuation area is added to the trapped zone. The numerical results from this modified algorithm are shown in the third line of Fig~\ref{fig:mosaic}. The initial trapped zones are relatively small and they appear as menisci of resident fluids. Near the breakthrough point, large zones of wetting fluid get trapped, and at the end of the simulation the saturation of the wetting fluid reaches a residual value. We compare the invasion pattern with the experimental results that are shown in the fourth line of Fig~\ref{fig:mosaic}. A perfect match is observed below the breakthrough point, and there are significant differences at the final stage. In the flow-controlled experiment, a viscous finger reaches the breakthought that produces a residual saturation larger than in the pressure-controlled simulation. This discrepancy is expected since the experiment is flow-controlled so that the capillary radius can increase during the injection. On the other hand, the simulations are pressure-controlled so that the capillary radius always decreases. The simulations are expected to reproduce the pressure-saturation relation when the pressure is quasistatically increased and the saturation is measured. The experiment includes dynamics effects such as viscosity that are not accounted in the simulations. However, it is reasonable to assume that viscous effects would be negligible in a quasi-static condition where the pressure is slowly increased and the saturation is recorded from the pressure-controlled experiments.

\section{Pore Morphology Method on square lattices}
\label{sec:simulations}

In the previous section, the void space occupied by both wetting (resident) and non-wetting (invading) fluids was calculated by quasi-static simulations of pressure-controlled experiments by increasing the capillary pressure, and hence decreasing the capillary radius and calculating the regions occupied by the wetting and non-wetting fluids. In this section, we calculate the saturation of these fluids, which are defined as the percentage of void space occupied for each fluid.  We focus on the variable $S$ that is the saturation of the resident (defensive) fluid, here refereed simply as {\it Saturation}.  The pressure-saturation relation depends on the material parameters $\gamma$ and $\theta$, as well as on the geometry and topology of the pore network of the void space. 

To investigate the dependency of the pressure-saturation relation on these parameters, we concentrate on the simplest pore network topology that is the square lattice.  Each pore in the square lattice is connected to its four neighbors by cylindrical throats. The radii throats are randomly generated with a log-normal distribution with median throat size $\tilde r$ and normal standard deviation $\sigma$. It is assumed that the contact angle $\theta$ is uniformly distributed in the range $\theta\pm\Delta\theta$. As boundary conditions, the injection zone is the top boundary of the rectangular sample, and the evacuation zone is the bottom boundary. 

Notice that the saturation will depend on the Pore Size Distribution (PSD) of the square lattice. More precisely, the saturation should be obtained by summing the area of all invaded pores, and dividing by the total void area. A simplified calculation of the saturation is given by the fraction of invaded pores. We observe that in the statistical limit, (i.e. for large enough samples or for averages over many samples), both calculations of saturation lead to similar results. Thus, the reported results in this paper used the simplified calculation of saturation as the fraction of invaded pores.

\subsection{Dimensional Analysis}
\label{subsec:dimensional analysis}

The parameters of our simulation are classified as material, statistical, and geometrical parameters. Our simulations are quasistatic, hence, the viscosity is absent, and the only material parameters are interfacial tension $\gamma$ and contact (wetting) angle $\theta$. In certain cases where the granulate is heterogeneous, the contact angle can vary from point to point. A simplified description of the material heterogeneities requires two parameters accounting for the mean value $\theta_0$ and the standard deviation $\Delta\theta$ of the contact angle. The statistical parameters describe the geometrical heterogeneity of the pore network. Being the pore throats log-normal distributed, we define $\tilde{r}$ as the median of the log-normal distribution and $\sigma$ the standard deviation of the normal distribution corresponding to the log-normal one. The geometric parameters are given by the averaged grain diameter $\bar{\ell}$, and the dimensions of the sample $L_x$ and $L_y$. The default parameters are listed in Table \ref{tab:Parameters}.

\begin{table}[t]
    \centering
    \begin{tabular}{c|c|c|c|c} 
        \hline
        Variable & Symbol & MS  & SL & Units \\ 
        \hline
        interfacial tension  & $\gamma$  & 55 & 50  & mN/m  \\ 
        mean contact angle   & $\theta_0$ &  60  & 50    & degrees\\
        contact angle dispersion     & $\Delta\theta$ & 0 & 0 & degrees\\
        median throat radius   & $\tilde{r}$  & -- & 10  & $\mu$m  \\ 
        $\sigma$-parameter log-normal& $\sigma $   & -- & 0.1   & -- \\ 
        average grain diameter       & $\bar{\ell}$    & 380 & 100 & $\mu$m \\
        sample length in x           & $L_x$        & 6 & 4  & mm  \\
        sample length in y           & $L_y$        & 4 & 4  & mm  \\
        graph/lattice                & g            & -- & square &--\\
        \hline
    \end{tabular}
    \caption{Parameters used in the Pore Morphology simulations. In the microchip simulation (MS), the porous geometry and material parameters are taken from the microchip experiment of Youtis et al. \cite{yiotis2021pore}. The fourth column is for the square-lattice (SL) simulations,  and it lists the default parameters of the simulation using random throat distributions.}
    \label{tab:Parameters}
\end{table}

The relation between the saturation of the defended (wetting) fluid $S$ and the capillary pressure $P_c$ in terms of the parameters of the model is 
\begin{equation}
    S = F_g(P_c,\gamma,\bar{r},\sigma,\theta,\Delta\theta,\bar{\ell},L_x,L_y),
\end{equation}
where $F_g$ is an unknown function for given contact network $g$.   Using dimensional analysis, it can be shown that the non-dimensional saturation variable will depend on dimensionless variables only:
\begin{equation}
    S = f_g(P_c\bar{r}/(\gamma\cos\theta_0),\sigma,\Delta\theta,L_x/\bar{R},L_y/L_x).
\end{equation}
From here we obtain the dimensionless pressure in terms of dimensionless quantities,
\begin{equation}
    \frac{P_c}{P_0}=\tilde{f}_g(S,\sigma,\Delta\theta,L_x/\bar{R},L_y/L_x),
    \label{eq:nondim}
\end{equation}
where,
\begin{equation}
    P_0 = \frac{\gamma\cos\theta_0}{\tilde{r}}.
 \label{eq:entry pressure}
\end{equation}
$P_0$ is the characteristic pressure, corresponding to the capillary entry pressure for a cylindrical tube of radius $\tilde{r}$.  Eq.~(\ref{eq:nondim}) defines the functional relation between the dimensionless capillary pressure and the saturation in terms of the dimensionless parameters $\sigma$ accounting for the variability of the pore throat radii, $\Delta\theta$ for heterogeneity of the material parameters of the granulate,  $L_x/\bar{R}$ for the size of the lattice, and  $L_y/L_x$ for its aspect ratio. 

\begin{figure}[b]
    \centering
    \includegraphics[trim={1cm 6.5cm 2cm 7.5cm},clip,width=\linewidth]{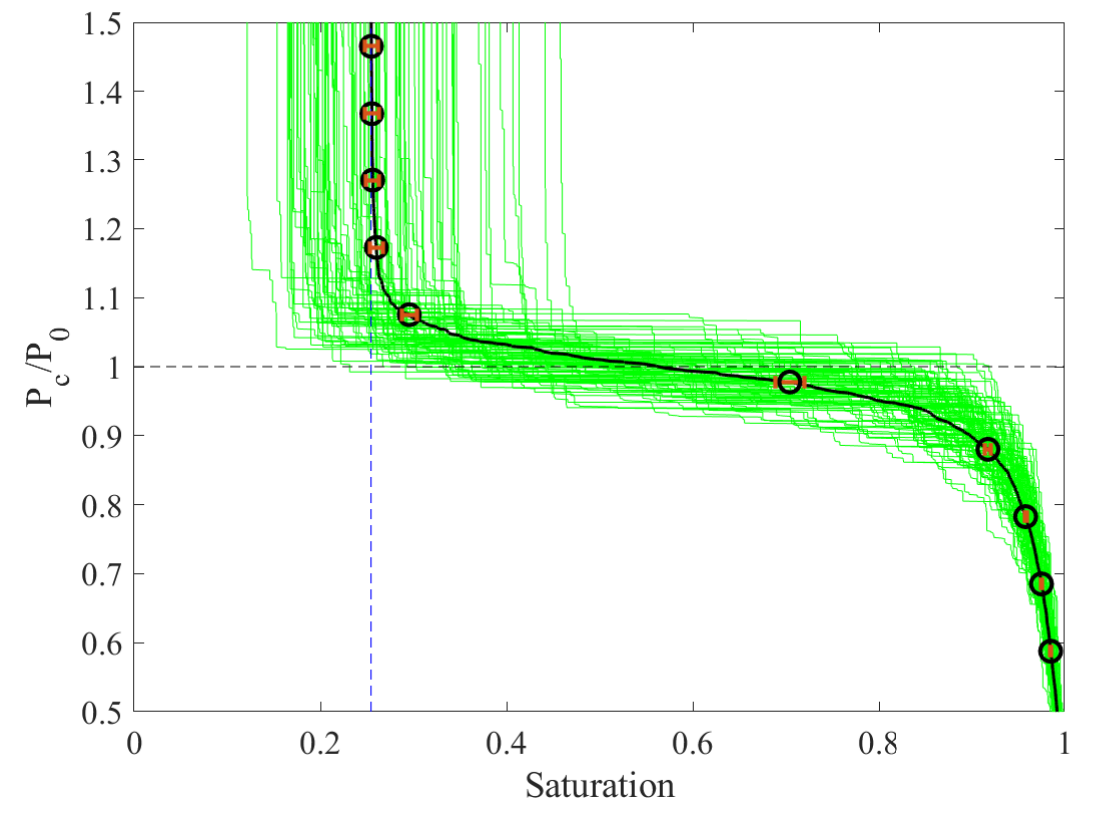}
    \caption{Dimensional capillary pressure versus saturation of the wetting (defended) fluid. The green lines are the result for individual samples, and the black one is the average over all samples. The dashed horizontal line represents the entry pressure where $P_c = P_0$, and the dashed vertical line is the residual saturation. The error bars are obtained by averaging the results over $96$ samples; the width of the error bars is the standard error.}
    \label{fig:PcVsS}
\end{figure}

\begin{figure*}
    \centering
    \includegraphics[trim={2cm 6cm 2cm 3cm},clip,width = 0.24\linewidth]{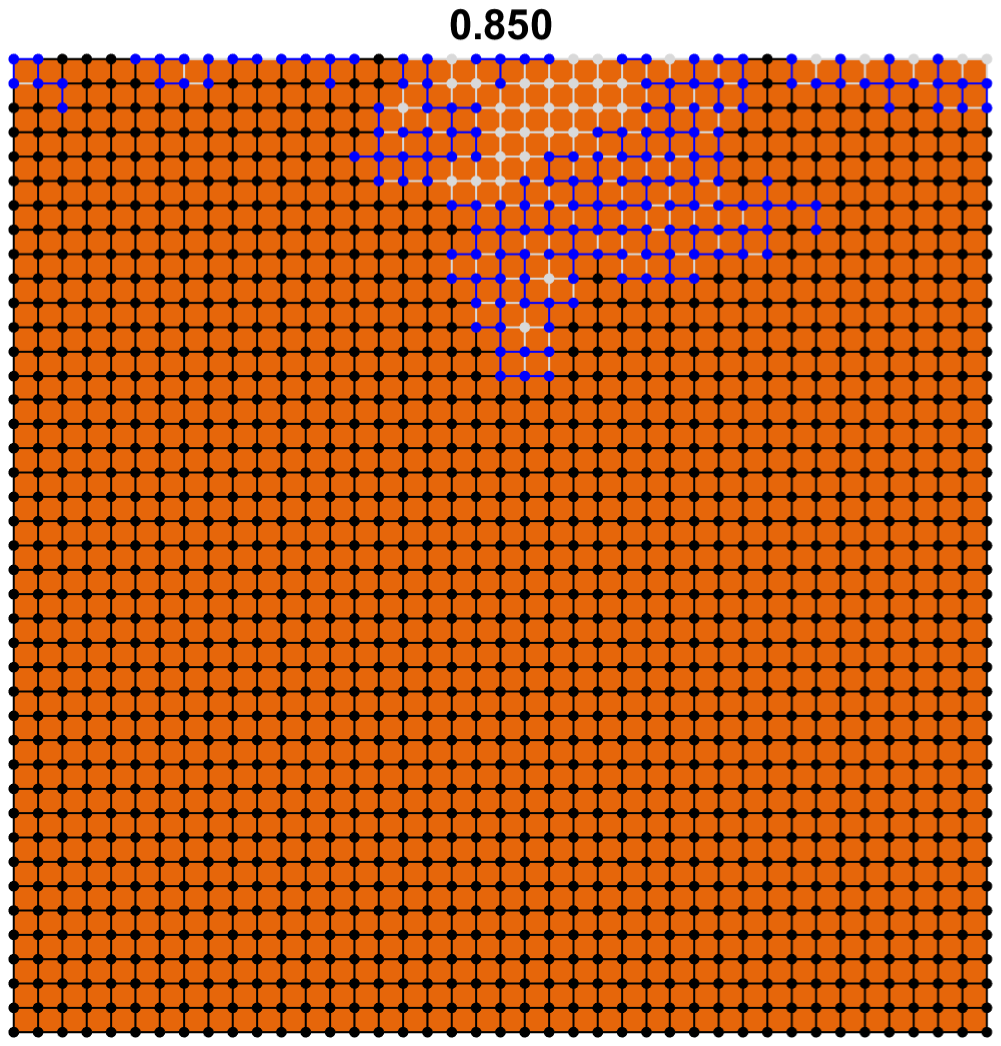}
    \includegraphics[trim={2cm 6cm 2cm 3cm},clip,width = 0.24\linewidth]{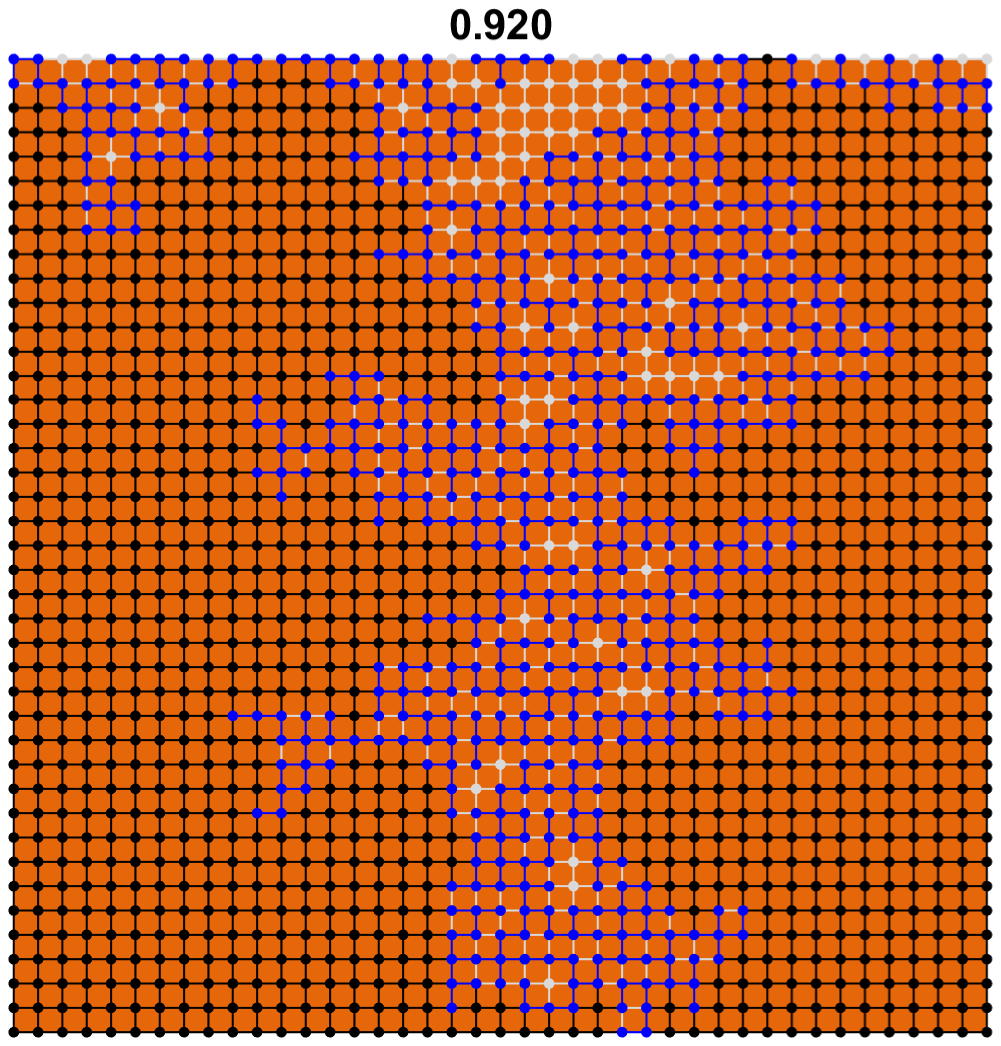}
    \includegraphics[trim={2cm 6cm 2cm 3cm},clip,width = 0.24\linewidth]{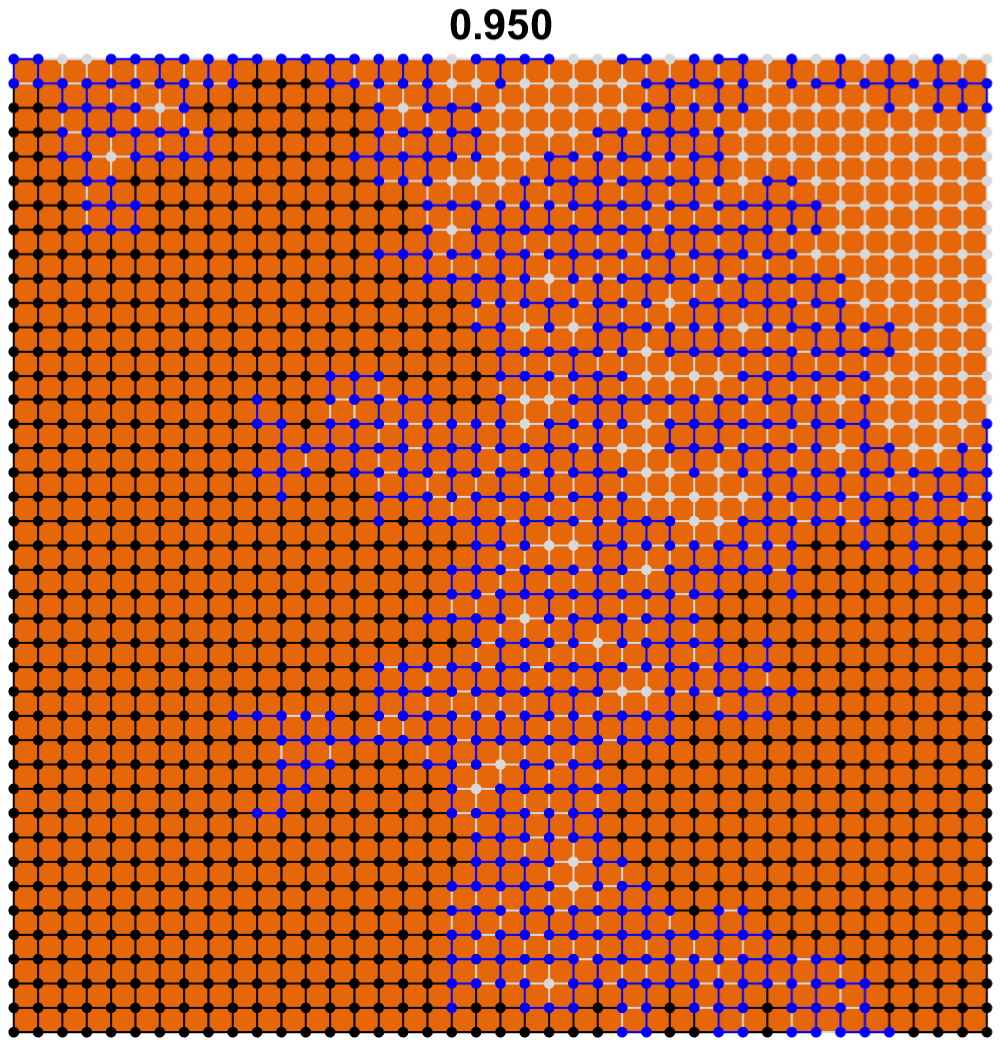}
    \includegraphics[trim={2cm 6cm 2cm 3cm},clip,width = 0.24\linewidth]{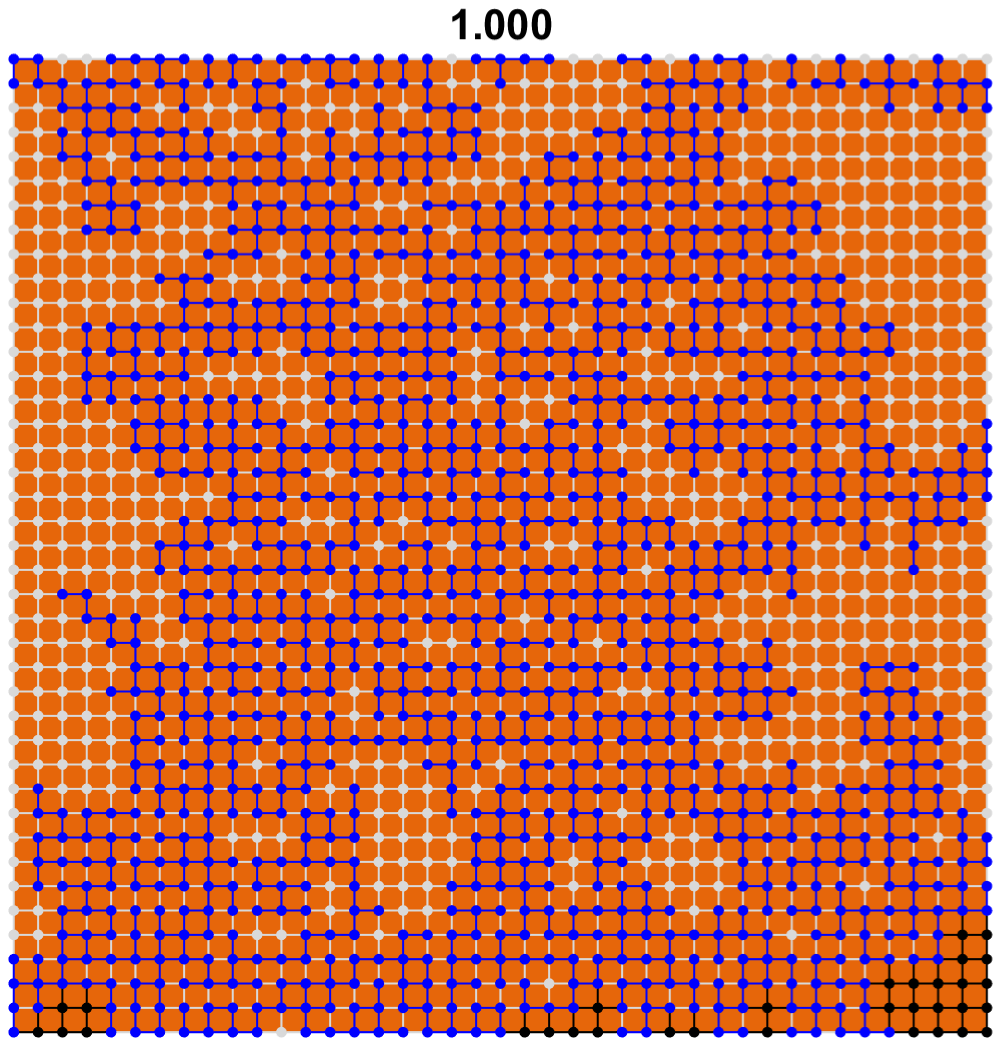}
    \caption{Snapshot of the pore invasion simulations for different values of $P_c/P_0$. the percolation transition occurs near $P_c = P_0$}
    \label{fig:invasionpercolation}
\end{figure*}

We investigate here the parametric dependency of the function $\tilde f$ on these dimensionless variables. 

\subsection{Pressure-Saturation relation}
\label{subsec:Pressure-saturation}

Here we calculate the relation between the dimensionless pressure and saturation for the default parameters listed in Table~\ref{tab:Parameters}. First, we generate random contact networks; each network has the same topology of the square lattice, but the pore throat distribution is randomly generated in each lattice. The results are shown in Fig. ~\ref{fig:PcVsS}. We observe strong statistical variation in the pressure-saturation relation. This variation is specially significant when the sample reaches the residual saturation. To obtain statistical representative results, we need to average the results over large number of samples. The averaged pressure-saturation relation shows an inflection point at $P= P_0$ that corresponds to the {\it entry pressure}.  Below the entry point, the saturation changes little with any increase of capillary pressure. On the other hand, when $P_c>P_0$ the saturation decreases with an increase of pressure, and it asymptotically reaches a residual saturation.  

The evolution of the pore invasion exhibits a behavior similar to other statistical processes such as bond percolation \cite{feng2008percolation}, backbone percolation \cite{sampaio2018elastic}, and invasion percolation \cite{wilkinson1983invasion}, see Fig.~\ref{fig:invasionpercolation}. The entry pressure $P_0$ resembles a critical percolation threshold, where the clusters of the invaded fluid spans over the size of the sample. A well-defined value near to the entry pressure is observed in each sample, where the breakthrough occurs and the saturation experiences a sudden decrease.  Slightly above this percolation threshold, the zone of invaded fluid produces a dense system of percolation paths, with resemblances to back-bone percolation threshold~\cite{sampaio2018elastic}. For large pressures, the invasion process reaches a steady state where the clusters of trapped fluid of all sizes are created. Also, the final pattern of trapped fluid is very sensitive to the initial seed of the random generation of the network. These features are typical of self-organized critical systems \cite{bak1988self}. It is evident in Fig.~\ref{fig:invasionpercolation} that large clusters appear on the boundaries, leading to boundary and finite-size effecst. These effects will be explored in Subsection~\ref{subsec:finite size}.

 \subsection{Effect of surface and/or interfacial tension}

\begin{figure}
    \centering
    \includegraphics[trim={6cm 6cm 5cm 7cm},width=0.5\linewidth]{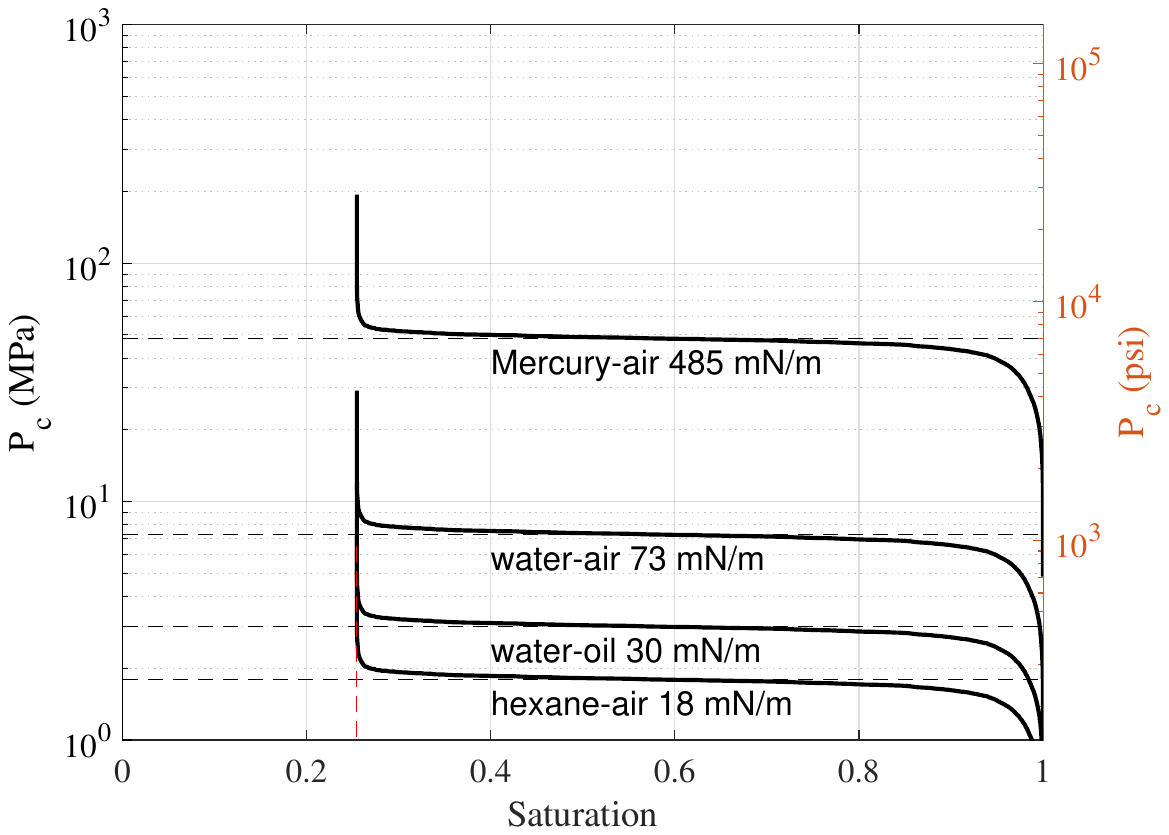}
    \caption{Pressure-saturation relation for different surface and interfacial tensions. The units of pressure are given in SI units (left) and Engineering PSI units (right).}
    \label{fig:SL_gamma}
\end{figure}

In virtue of the dimensional analysis, the pressure-saturation relation in Fig ~\ref{fig:PcVsS} is the same for arbitrary interfacial tensions $\gamma$ and contact angles $\theta_0$. We can recover the dimensionality by selecting these material parameters and plotting the dimensional capillary pressure against saturation. Figure \ref{fig:SL_gamma} shows the curves for typical surface tension between mercury, air, water, oil, and hexane. An increase of surface tension leads to an increase of the entry pressure while all pressure-saturation curves are self-similar, and they share the same residual saturation, i.e. the residual saturation does not depend on the surface tension. Let us notice that residual saturation may be affected by surface and interfacial tension in a real experiment. Our simulations just show that in the special case of pressure-controlled and quasistatic pore invasion in immiscible fluids, surface tension does not play a role in the residual saturation.

\subsection{Effect of heterogeneity of pore network}

\begin{figure*}
    \centering
    \includegraphics[trim={1cm 11cm 2cm 11cm},clip,width=\linewidth]{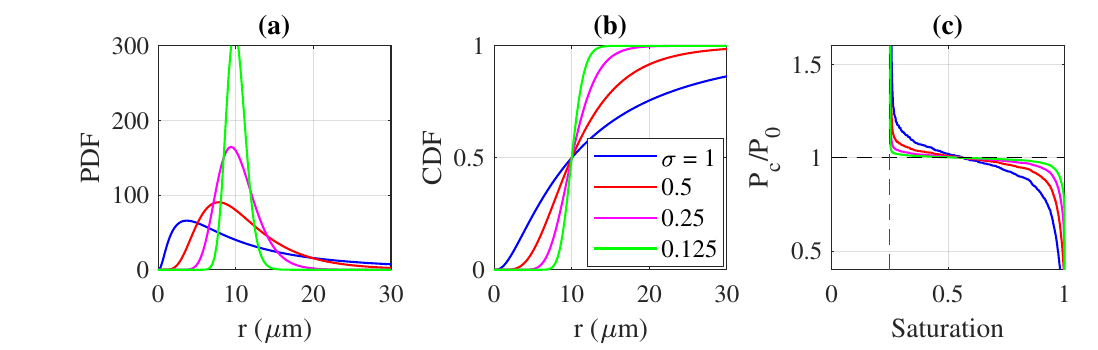}
    \caption{(a) Different variances in the pore throat distributions with the same media $\tilde{r}= 10\mu m$. (b) Cummulative distribution of throat distributions. (c) Pressure-saturation relation for the different throat variances share the same entry pressure and residual saturations (dotted lines).}
    \label{fig:SL_sigma}
\end{figure*}

For a given pore topology such as in our case of the square lattice, the statistical characterization is given by the Pore Size Distribution (PSD) and Pore Throat Distribution (PTS). The PSD directly affects residual saturation, but in the statistical limit the saturation can be given in terms of the fraction of the invaded pores. The PTS affects the pressure-saturation relation in two ways, the median of PTS  $\tilde{r}$ is inversely proportional to the entry pressure, as shown in Eq.~(\ref{eq:entry pressure}). The parameter of the standard deviation $\sigma$ affect the pressure-saturation relation via Eq~(\ref{eq:nondim}). Here we investigate the effect of the variance of the PTD.

Again, we can use the dimensional analysis to investigate the dependency of the pressure-saturation relation on the broadness of the pore throat distribution (PTD) while fixing the other parameters. For a given parameter $\tilde r$ measuring the median of the pore throat distribution, the parameter $P_0$ is calculated using Eq.~(\ref{eq:entry pressure}). Then the relation between the dimensionless pressure $P_c/P_0$ and saturation is calculated using Eq.~(\ref{eq:nondim}). The results are shown in Fig.~\ref{fig:SL_sigma}. The median of the PDF is fixed to $\tilde r = 10 \mu m$ and the variance $\sigma$ is changed. For all values of $\sigma$, the pressure-saturation relations share the same entry pressure and residual saturation. The effect of $\sigma$ is on the sharpness of the transition around the entry pressure: if $\sigma$ is small (narrow PTD), the transition is sharp and becomes smoother as $\sigma$ increases (wider PTD).

\subsection{Effect of heterogeneity of the granulate}
\label{subsec:hetero}
\begin{figure}
    \centering
    \includegraphics[trim={6cm 6cm 6cm 6cm},width=0.5\linewidth]{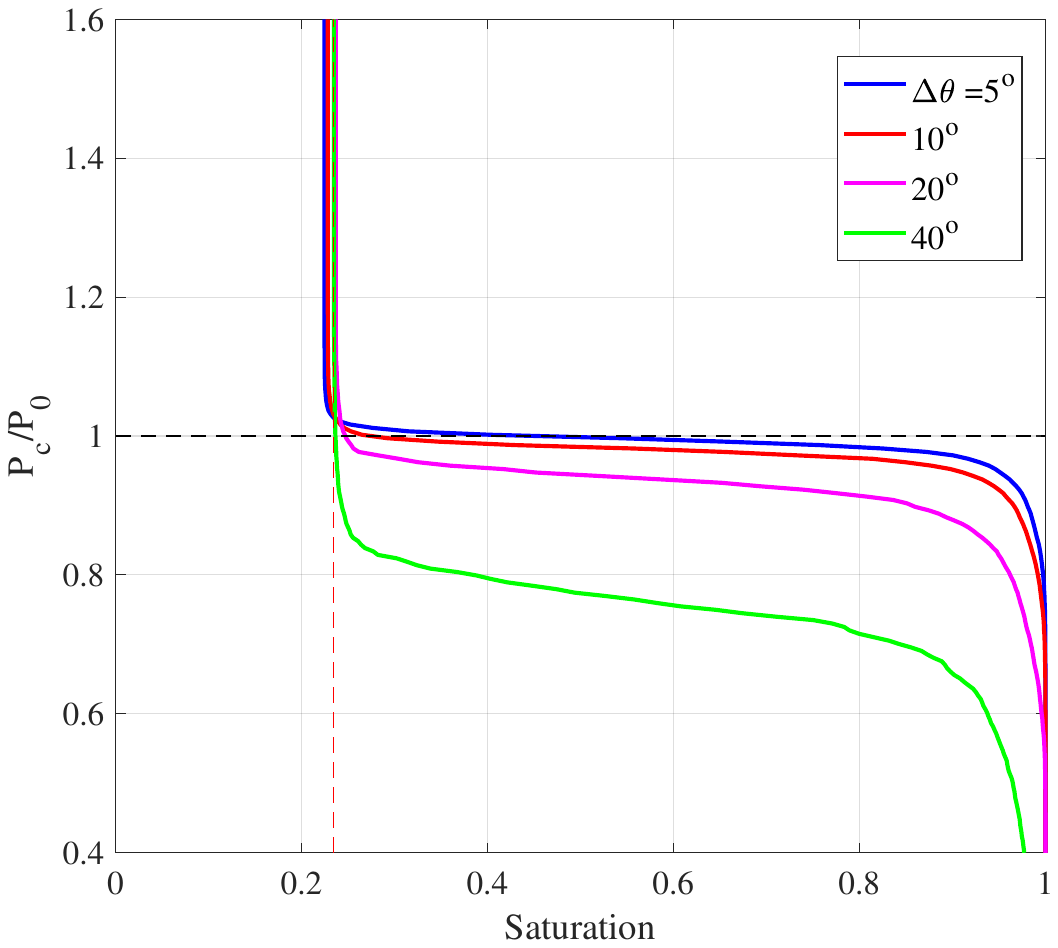}
    \caption{Pressure-saturation relation with different variability of the contact angle. The curves does not share the same entry pressure, and it vanished in the extreme case when $\Delta\theta = 40^0$. This is the case when the contact angle changes from $10^0$ to $90^0$. All curves share the same residual saturation.}
    \label{fig:SL_deltatheta}
\end{figure}

The previous results accounts for the geometrical heterogeneity of the pore network. It is also relevant to investigate how the pressure-saturation relation depends on how heterogeneous the granular matrix is. Natural porous media such as sands and sandstones have different mineral compositions, typically including quartz, felsdpar, heavy minerals, and cementing materials \cite{boggs2006principles}. Carbonate rocks are composed mainly of calcite and dolomite \cite{sperber1984rock}. The mineral composition directly affects the wettability, so that the contact angle $\theta$ is expected to vary from point to point on the granulate. Material heterogeneity is accounted by the parameter $\Delta\theta$ that indicates how broad the variance of the contact angle is distributed. The contact angle is uniformly changed in the range $\theta_0 \pm \Delta\theta$. Fig. ~\ref{fig:SL_deltatheta} shows the dependency of the pressure-saturation relation on the dispersion of $\theta$. We note that an increase of variance of the contact angle distribution produces a decrease of the entry pressure and a smoother transition around the entry pressure. This comes from the lower values of $\theta$ present in the wider distribution, with leads to a lower entry pressure, Fig.~\ref{fig:SL_gamma}. The actual shape of the pressure-saturation curve does depend on the distribution and for more quantitative results, a more realistic representation of the contact angle distribution would be warranted. Interestingly, all curves share the same value of residual saturation, indicating that, again, the residual saturation does not depend on the variability of the contact angle.

\subsection{Finite-size effect}
\label{subsec:finite size}
The results shown in the previous section were performed by a fixed sample dimension $4 mm \times 4 mm $ with a grain diameter of $0.1 mm$. Here we address the question on whether this sample is representative enough for using the pressure-saturation relation for large-scale simulations. In other words, we ask: How is the pressure-saturation relation affected by an increase of sample size? For this analysis, we fix the aspect ratio of the sample to one and change the sample dimensions. All other parameters are fixed.

\begin{figure}
    \centering
    \includegraphics[trim={6cm 6cm 6cm 6cm},width=0.5\linewidth]{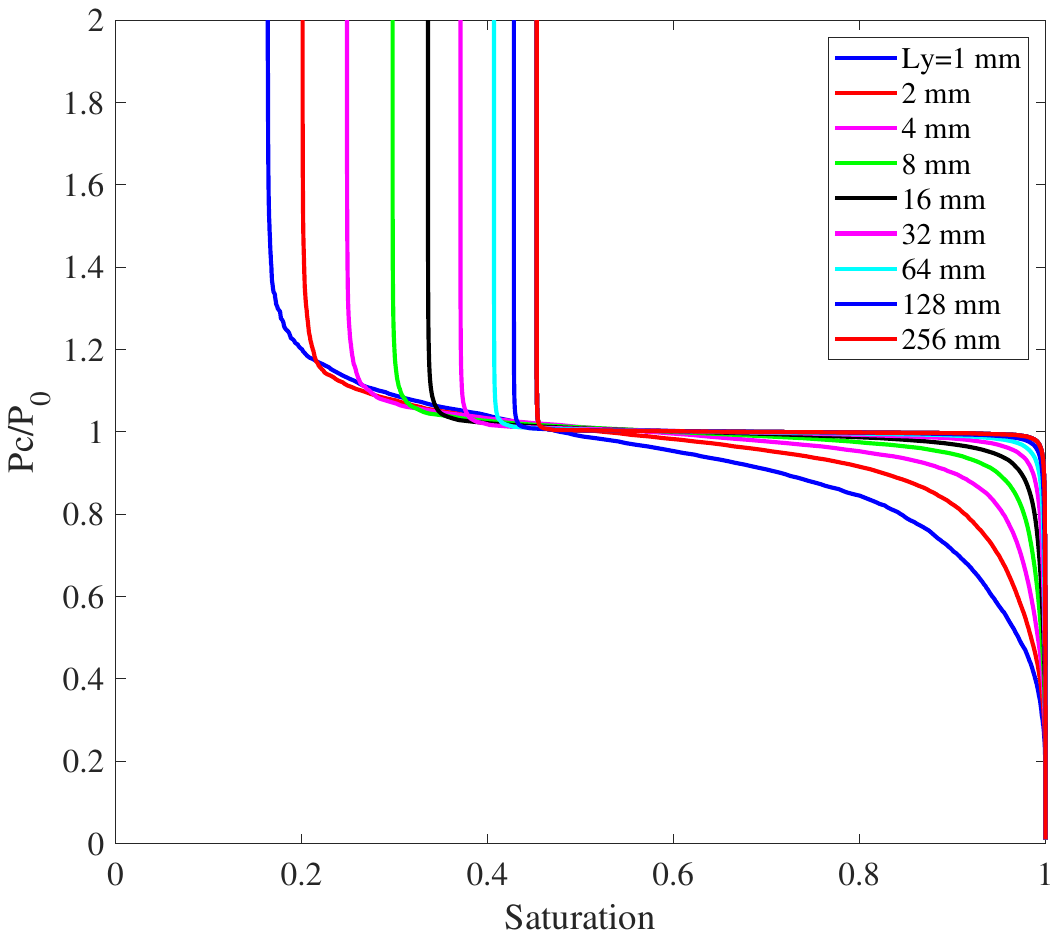}
    \includegraphics[trim={6cm 6cm 6cm 6cm},width=0.5\linewidth]{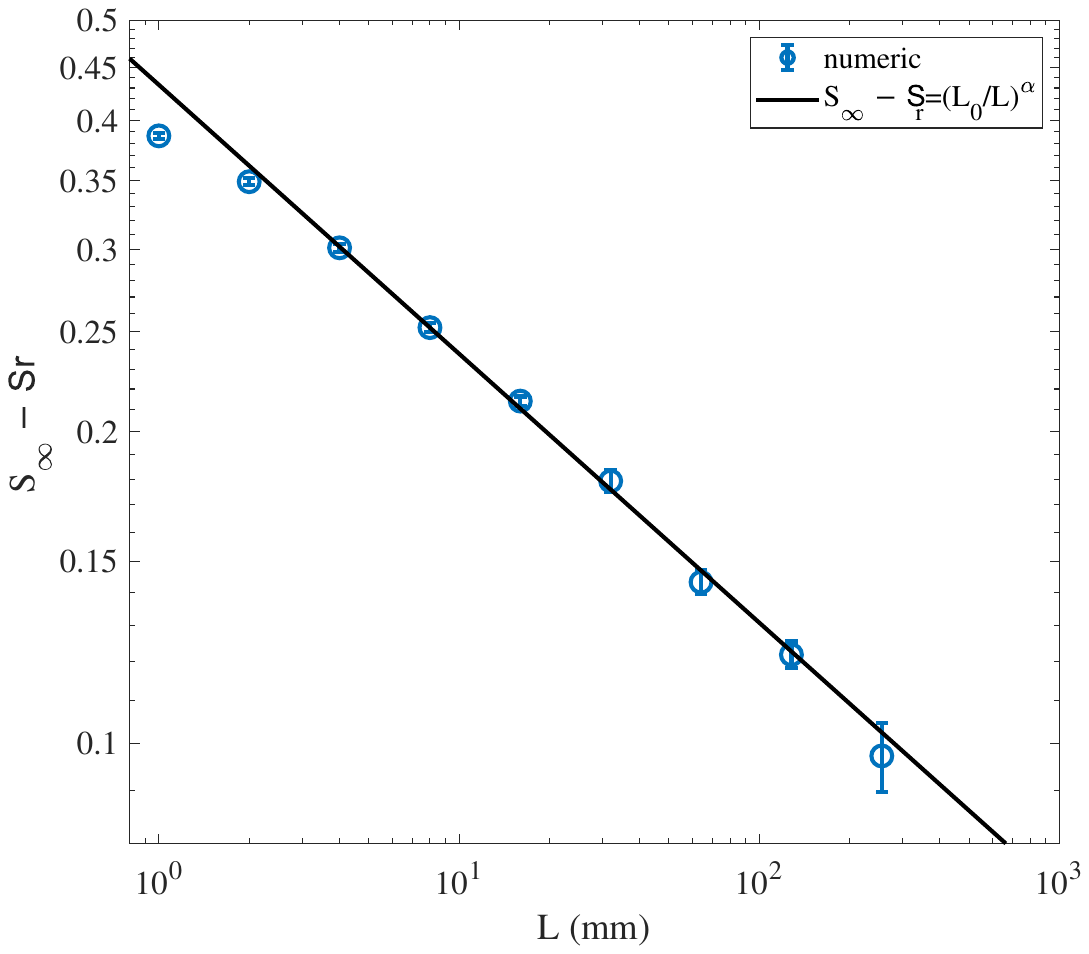}
    \caption{(a) Pressure-saturation relationship for different samples size on square samples. (b) Dependency of the residual saturation on the sample size, The residual saturation is fitted by the relationship $S_r =S_\infty -(L_0/L)^\alpha$, where $S_\infty =0.55 \pm 0.02$, $L_0=(0.04\pm 0.005)$ mm, and $\alpha = 0.26 \pm 0.02$. }
    \label{fig:SL_samplesize}
\end{figure}

The dependency of the pressure-saturation relation on sample size is shown in Fig.~\ref{fig:SL_samplesize}a. The entry pressure does not change with the sample size, but as the sample size increases, the transition around the entry pressure becomes sharper. The main effect of changing the size of the sample is the change of the residual saturation; larger samples lead to higher residual saturation. It is also remarkable how slow the convergence of the residual saturation is, as the sample size increases. Simulations with samples of $0.26 m$ were performed, and we still did not achieve convergence of the residual saturation.

To estimate the residual saturation for infinite samples, and to verify convergence, we fit the dependency of residual saturation $S_r$ to a power law. The fitting parameters is the residual saturation for infinite samples $S_\infty$ and the power-law exponent $\alpha$, i. e. $S_r=S_\infty -(L_0/L)^\alpha$. The result of the power-law fit is show in Fig ~\ref{fig:SL_samplesize}b. We achieve a very good fitting except for very small samples where boundary effect may produce deviations from the power law. The exponent of the power law is $0.26$ indicating a very slow convergency of the residual saturation. The residual saturation for infinite samples is quite high, $ 55\% $. It is expected that this is the characteristic value for square lattices, and it should depend on the topology of the lattice and its dimensionality.

\begin{figure*}
    \centering
    \includegraphics[trim={2cm 4cm 2cm 4cm},clip,width = 0.4\linewidth]{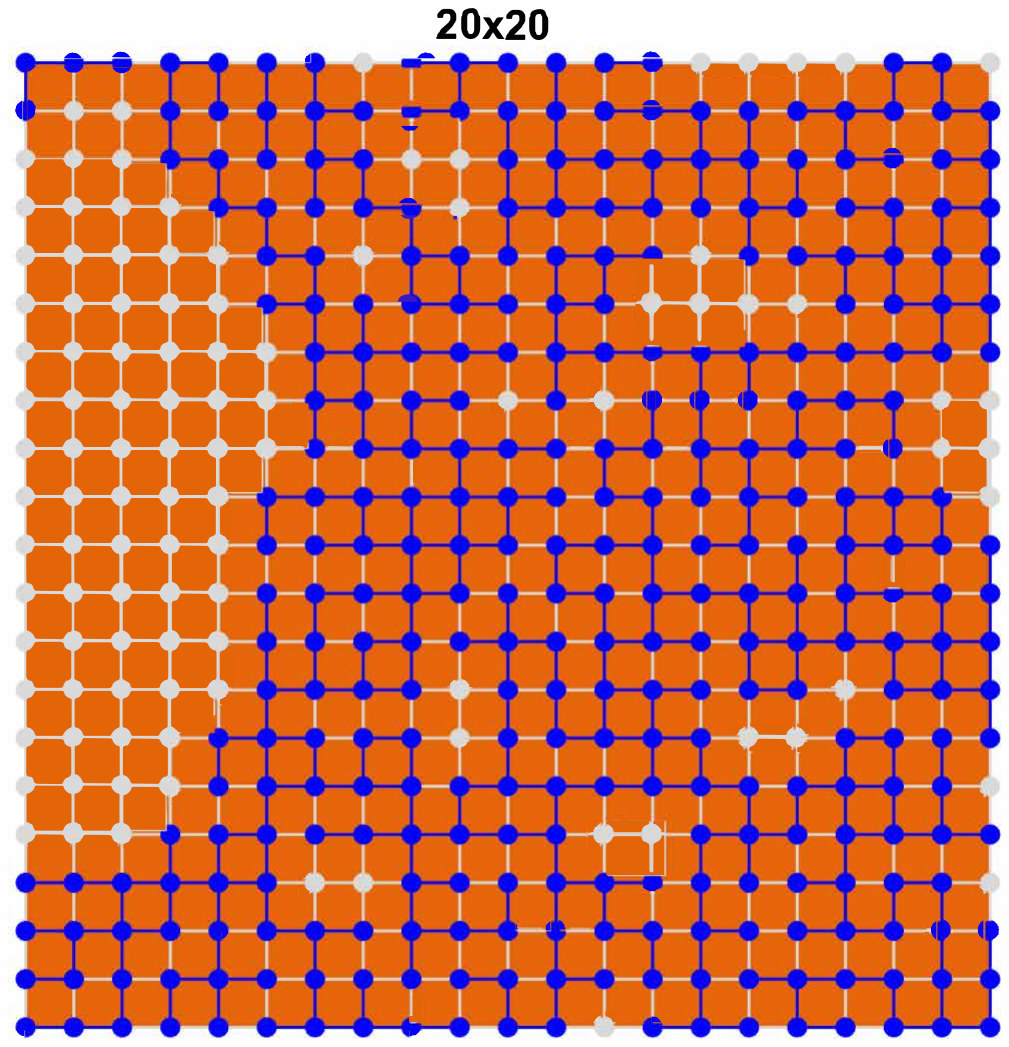}
    \includegraphics[trim={2cm 4cm 2cm 4cm},clip,width = 0.4\linewidth]{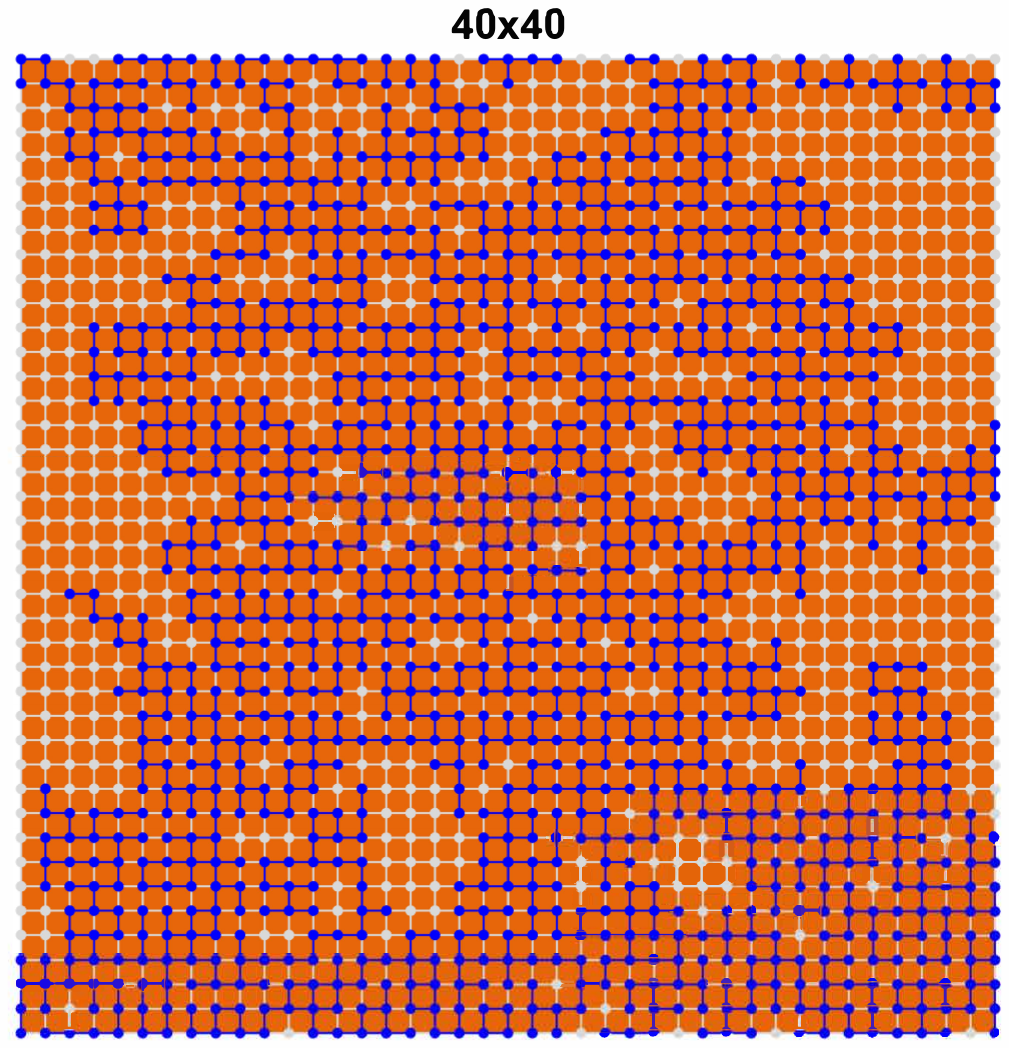}\\
     \includegraphics[trim={2cm 4cm 2cm 4cm},clip,width = 0.4\linewidth]{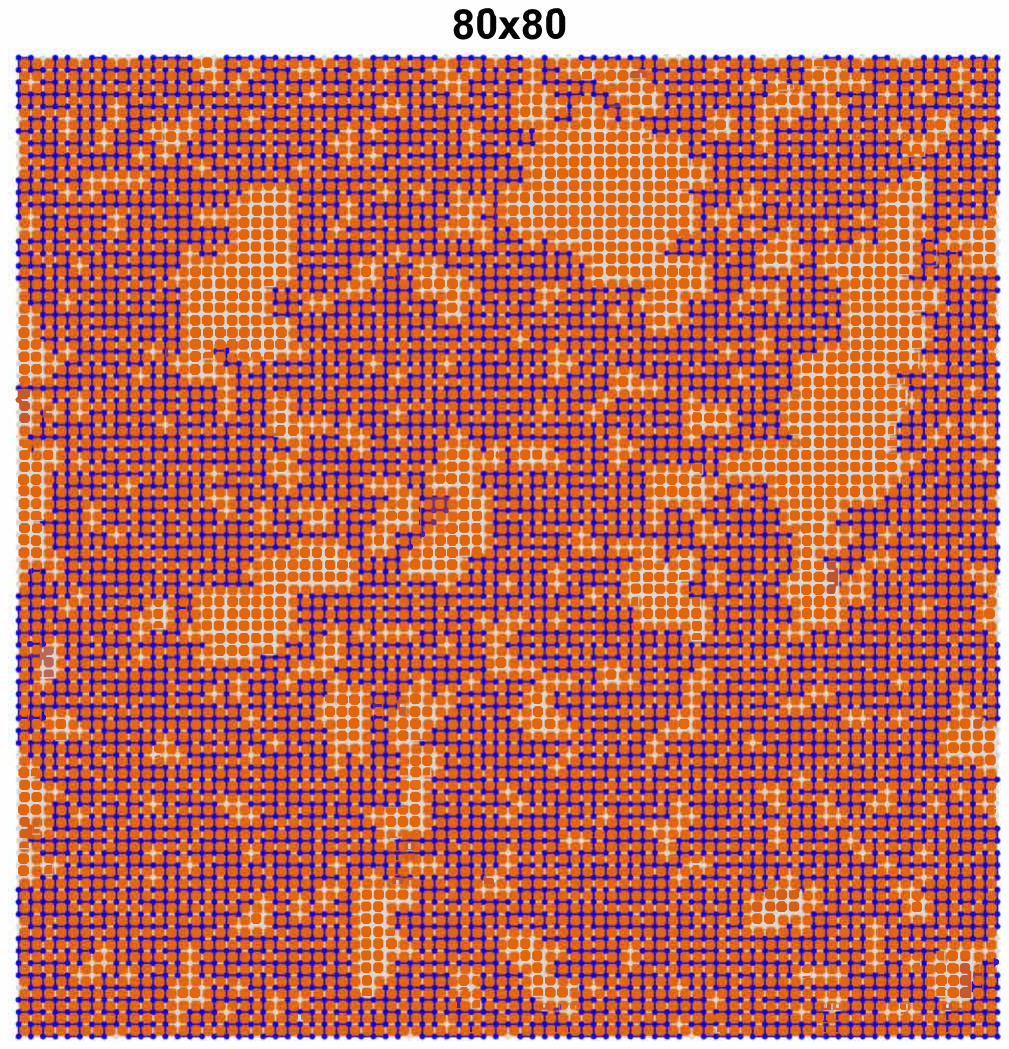}
      \includegraphics[trim={2cm 4cm 2cm 4cm},clip,width = 0.4\linewidth]{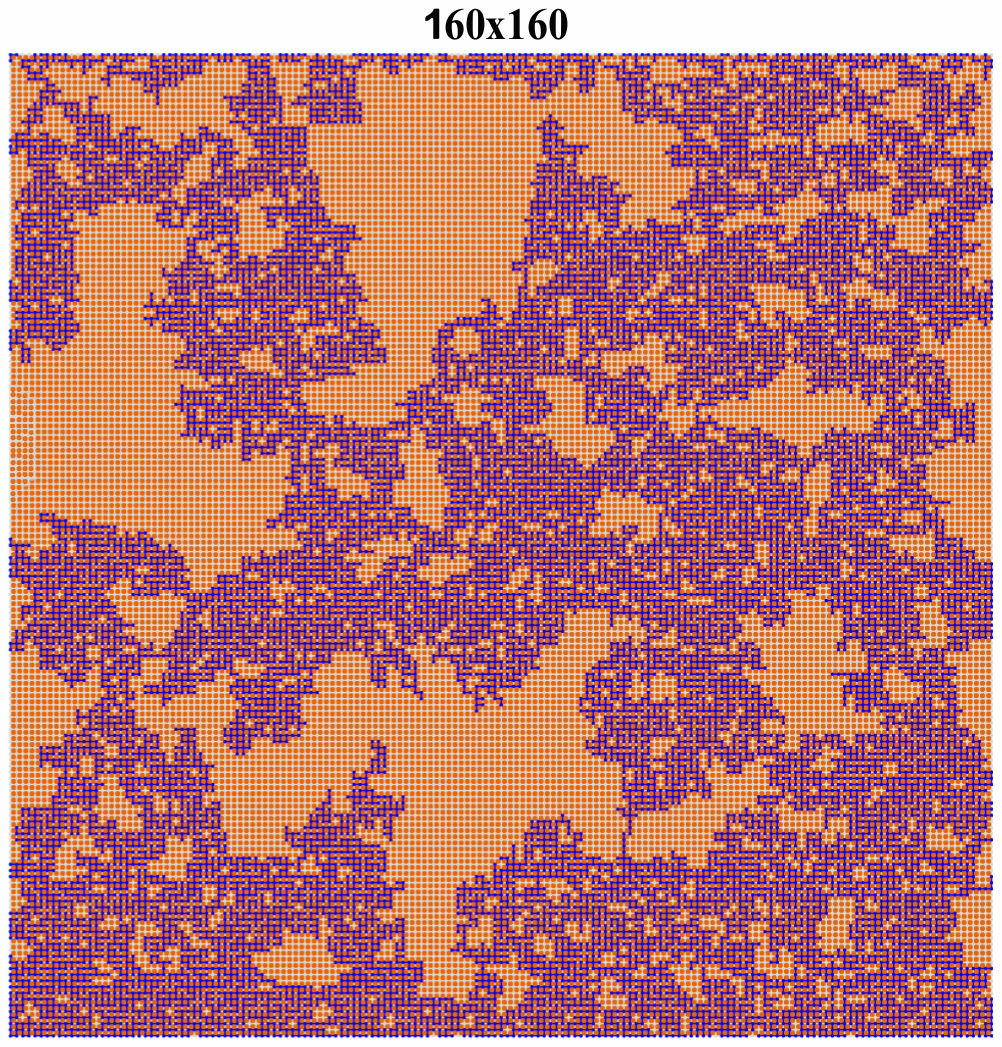}
      \label{fig:sizes}
      \caption{Final distribution of resident (gray) and invaded (blue) zones for different sample sizes. The dominant clusters of trapped fluid are larger for larger samples, a phenomenon that is characteristic of self-organized criticality systems. }
\end{figure*}

To understand the effect of sample size on residual saturation, we plot the final distribution of fluids for different sample sizes, see Fig.~\ref{fig:sizes}. for samples below $0.4 mm$ size length, the boundary effects are evident, large clusters of trapped fluids appear mostly on the boundary of the fluid. As the sample is enlarged, the relation between perimeter and area decreases, and hence the boundary effects are reduced. However, it is remarkable that the cluster size keeps increasing as the sample size increases. Typically, the size of the largest cluster is of the order of the sample size. There is no apparent averaged cluster size. This is probably the reason for the ultra-slow convergence of the residual saturation as the sample size increases.

\section{Discussion of the results}
\label{sec:discussion}

We have presented the results of quasistatic simulations of two immiscible fluids, one displacing the other, in a porous medium. The control parameter of the simulation is the capillary radius: as the capillary pressure is increased, the capillary radius decreases, allowing the invaded fluids to access more porous space, and hence, decreasing the saturation of the resident fluid. We have identified two variables to characterize the pressure saturation relation: the dimensionless entry pressure and the residual saturation. The only parameter affecting the dimensionless entry pressure is the variance of the contact angle, but it is expected that the entry pressure will change for different pore-network topologies, apart from the square lattice used in these calculations.  The residual saturation seems to be unaffected neither the pore throat distribution nor the material properties.

Residual saturation depends only on sample size, and it is expected to depend on the topology of the porous network that in our case is a simple square lattice. Residual saturation depends neither on material parameters (interfacial tension and wettability) nor on the pore throat distribution (median and standard deviation of pore throat distribution). This is in apparent contradiction to many experimental results showing that decreasing interfacial tension and wettability reduces residual saturation. These discrepancies are due to the way the invasion is performed. In most experiments, including the one we compared our results to in this paper ~\cite{yiotis2021pore}, the fluid invasion is made using a flow-controlled experiment, while our simulations are pressure-controlled. On a flow-controlled experiment, the pressure drops each time the invaded fluid breaks through a throat, leading to an increase of the capillary radius, in virtue to Eq.~\ref{eq:capillary radius}. This is evident in Fig~\ref{fig:mosaic} where the residual saturation in the pressure-controlled test is lower than the one in the flow-controlled one.  Thus, these pressure drops reduced the space the non-wetting fluid can invade with respect to the hypothetical case where the pressure is not allowed to drop (and hence, the capillary radius does not increase). The residual saturation presented here should be interpreted as the lower bound of all possible saturation values achieved during the flow-controlled experiments.

We also should notice that the limit of the pressure-saturation curve as the sample size increases is particularly restrictive to a step-like function defined by the entry pressure and residual saturation. This step function rules out typical pressure-saturation relations observed in the literature. In particular, experiments on sandstone samples do not show a plateau in the pressure-saturation relation, but a semi-logarithmic relation between pressure and saturation instead \cite{cao2016pore}. We should notice that these discrepancies are expected since we focus on the simplest pore network given by the square lattice. The pore network topology in sandstone is much more intricate, often exhibiting multiple scales and spacial correlations that are not accounted here. A more general description of the pressure-saturation relation will require more diverse pore network topologies and spatial correlation of both materials and geometrical heterogeneities.

We also point out that our simulations are quasistatic and two-dimensional, which impose restriction when compared to real scenarios. In principle, the Pore Morphology Method can be extended to 3D simulations, but it will require more computational demanding calculations, Also, it is possible to add dynamic effects on the Pore Morphology Methods by imposing evolution rules on the dynamics of the capillary radius during the invasion process. This extension will provide an interesting numerical counterpart to advanced voxel-dynamics simulations. Our interest in this paper was on quasistatic simulations that provides a computationally efficient calculation of the pressure-saturation relation that is under-explored in the literature.

\section{Conclusions}
\label{sec:conclusions}
In this paper, we have presented an extension of the original Hilper and Miller algorithm to calculate the relation between the capillary pressure and saturation in multiphase flow of immiscible fluids in porous media. The extension includes  effects of wettability and trapped mechanism, not included in the original formulation. The material parameters are given in terms of surface energies that are calculated using the Density Functional Theory. Based on dimensional analysis, we achieved excellent agreement with experimental results, and presented a comprehensive study of the effect of pore throat distribution, material heterogeneity, and sample size. The numerical results were restricted to the simplified topology of the square lattice used in the simulations, but they can be extended to more general topologies. 

The most salient result is the emerging criticality that results from the finite-size effects. This criticality is reflected into an ultra-slow convergency of the residual saturation due to the emergence of clusters of trapped fluids of all sizes, a phenomenon that is typical of Percolation Theory. An extended framework for the pore-scale flow will require the use of the Percolation Theory to identify universal exponents in the pressure-saturation relationship, universality classes for different pore network topologies, and self-similar profiles within these universality classes. Our extended Pore Morphology Method would provide a computational efficient numerical framework to build a comprehensive Percolation Model for multiphase flow for a wide range of applications.

\begin{acknowledgments}
FAM acknowledges useful discussions within the Integrative-Multiscale Modeling group and with Morteza N. Najafi. 
\end{acknowledgments}

\bibliography{main}

\end{document}